\documentclass[doublecol]{epl2} 
\newcommand{\bra}[1]{\left\langle#1\right|}
\newcommand{\ket}[1]{\left|#1\right\rangle}

\title{Combining the Hybrid Functional Method\\ with Dynamical Mean-Field Theory}
\author{D. Jacob \and K. Haule \and G. Kotliar}
\shortauthor{D. Jacob \etal}

\institute{                    
  Dept. of Physics \& Astronomy, Rutgers University, 136 Frelinghuysen Road, Piscataway, NJ-08854, USA
}


\abstract{
  We present a new method to compute the electronic structure of correlated materials
  combining the hybrid functional method with the dynamical mean-field theory.
  As a test example of the method we study cerium sesquioxide, a strongly correlated
  Mott-band insulator. The hybrid functional part improves the magnitude of the $pd$-band 
  gap which is underestimated in the standard approximations to density functional 
  theory while the dynamical mean-field theory part splits the $4f$-electron spectra 
  into a lower and an upper Hubbard band.
}

\begin{document}

\maketitle

\section{Introduction}
Recently, there has been considerable progress in the realistic description of strongly 
correlated materials by combining density functional theory \cite{Hohenberg:pr:64} 
(DFT) with the dynamical mean-field theory (DMFT)
\cite{Metzner:prl:89,Georges:rmp:96,Held:psik:03,Kotliar:rmp:06}. In this DFT+DMFT 
approach \cite{Anisimov:jpcm:97b,Lichtenstein:prb:98}, DFT 
is employed to obtain an effective
mean-field description of the weakly correlated bands while the local correlations in the 
more strongly correlated bands ({\it i.e.} the $d$-bands of transition metals and the $f$-bands of 
Lanthanides and Actinides) are treated exactly. The DFT+DMFT method allows to 
predict spectra and energies of stronlgly correlated materials. 

The DFT+DMFT method has been successfully applied to a variety of interesting materials 
that conventional band structure theory is unable to deal with. For example,  using the
DFT+DMFT method the 25\% of volume increase in the transition from the $\alpha$- to the 
$\delta$-phase of Pu could be explained by the presence of strong correlations in 
$\delta$-Pu\cite{Savrasov:nature:01}. 
However, by construction the DFT+DMFT approach, does not work so well in situations 
where the one-electron spectra of the weakly correlated bands are not well approximated 
by the Kohn-Sham (KS) spectra of DFT. For example, the local density approximation (LDA)
and the generalized gradient approximation (GGA) notoriously underestimate 
the magnitudes of band gaps of insulating materials. On the other hand within the 
chemistry community, very accurate functionals called hybrid functionals 
\cite{Becke:jcp:93a}, have been constructed by mixing LDA/GGA functionals with 
Hartree-Fock. The hybrid functional (HYF) approach has been tremendously successful in providing 
very accurate energies for molecules. Moreover, one-electron spectra computed with 
HYFs give fairly accurate gaps for semiconducting materials \cite{Muscat:cpl:01}.

In this work we propose a new method that combines the HYF approach 
with DMFT (HYF+DMFT) to yield a quantitatively \emph{and} qualitatively
correct description of combined band and Mott-Hubbard insulators. 
The HYF part improves the effective static mean-field description of 
the uncorrelated electrons while the DMFT part describes the dynamical 
local electronic correlations of the strongly localized electrons 
that can lead to Mott insulating behaviour. 
Although the HYF approach
introduces a new parameter, $\alpha$ that determines the amount of 
Hartree-Fock exchange, we show that this $\alpha$ can actually 
be linked to the Coulomb repulsion parameter $U$ of the DMFT 
calculation. 

An important example which illustrates the need for the HYF+DMFT method is provided
by the rare earth sesquioxides \cite{Adachi:chemrev:98} series which are
insulators. In addition to the Mott-Hubbard gap between the occupied and the 
unoccupied $4f$-bands, a band gap between the uncorrelated O $2p$- and $5d$-bands opens. 
One thus has to deal with the two-fold problem of finding an accurate description for the 
$pd$-band gap  and the $4f$-Hubbard bands in the same material. 

DFT and related static mean-field methods fail to describe the splitting of the $4f$
Hubbard band, without invoking some form of magnetic long range order. For example,
HYFs give the correct magnitude for the band gap of antiferromagnetic 
(AF) Ce$_2$O$_3$\cite{Hay:jcp:06,DaSilva:prb:07}. However, the symmetry-breaking is essential 
to obtain the correct insulating behaviour although it is clear that the magnetism should 
not be the driving mechanism behind the insulating behaviour since the N\'eel temperature 
is only 9~K and thus much smaller than the measured gap of about 2.5~eV. Similarly, 
symmetry-breaking is crucial in order to capture the insulating behaviour of Ce$_2$O$_3$ 
and of Ce$_2$S$_3$ with the DFT+U method \cite{Anisimov:prb:91,Andersson:prb:07,Loschen:prb:07,Windiks:08}.
Thus a satisfactory description of Ce$_2$O$_3$ should also yield the correct insulating behaviour 
in the paramagnetic phase. 

Correlations on top of DFT, can be added within the DFT+DMFT method. Indeed recent studies of the rare 
earth oxides have been carried out within this approach. The DFT+DMFT \cite{Pourovskii:prb:07} 
approach, succesfully describes the opening of the Hubbard band in the $4f$-shell of these 
materials. However, they underestimate the $pd$-band gap. It is natural then to combine 
the virtues of the HYF and the DMFT method, in a HYF+DMFT approach, which is the subject 
of this paper.

\section{Method}
The HYF+DMFT method follows the well established DFT+DMFT
methodology. We focus on the one-particle Green's function: 
\begin{equation}
  \label{eq:Gk}
  \hat{G}({\bf k},\omega) = 
  (\omega+\mu-\hat{H}({\bf k})-\hat\Sigma(\omega))^{-1}
\end{equation}
which is expressed in terms of a one-body Hamiltonian $\hat{H}({\bf k})$ 
and a local self-energy $\hat\Sigma(\omega)$. The self-energy describes the 
dynamic electron correlations of the strongly localized electrons 
and thus has non-zero elements only in the block of correlated bands. 
The correlations are captured by a Hubbard-like term which is added to the 
one-body Hamiltonian $\hat{H}({\bf k})$ in the strongly-correlated subspace:
\begin{equation}
  \label{eq:hubbard}
  \hat{\mathcal H}_U = \frac{1}{2}\sum_{a_1,b_1,\sigma_1 \atop a_2,b_2,\sigma_2} U_{a_1a_2b_1b_2} \,
  \hat{c}_{a_1\sigma_1}^\dagger\hat{c}_{a_2\sigma_2}^\dagger \hat{c}_{b_2\sigma_2}\hat{c}_{b_1\sigma_1}
 \end{equation}
where the indices $a_1,a_2,b_1,b_2$ denote orbitals of the correlated 
subspace in a local basis set.

Within DFT+DMFT the one-body Hamiltonian $\hat{H}({\bf k})$ is given by the 
effective KS Hamiltonian $\hat{H}_{\rm ks}=-\frac{\hbar^2}{2m}\nabla^2+v_{\rm ext}({\bf r})+
v_{\rm h}({\bf r})+v_{\rm xc}({\bf r})$. The KS Hamiltonian gives an effective one-body 
description of the electronic structure taking into account the Coulomb 
interaction on a mean-field level by the Hartree potential $v_{\rm h}$ 
and the exchange-correlation (XC) potential $v_{\rm xc}$. 

One of the main shortcomings of the standard approximations to DFT like LDA and 
GGA is their difficulty to describe insulating materials. On the one hand the 
insufficient cancellation of the self-interaction error by the approximate 
LDA and GGA XC functionals results in band gaps that are generally too small 
compared to the measured band gaps in semiconductors and insulators 
\cite{Muscat:cpl:01}. On the other hand DFT is strictly speaking only a ground 
state theory. Thus it does not necessarily give a correct description of excited 
state properties like the band gap of insulating materials. 

The HYF method \cite{Becke:jcp:93a} improves on LDA/GGA
by introducing a fraction $\alpha$ of exact Hartree-Fock exchange
into the XC potential:
\begin{equation} 
  \hat{v}^{\rm hyf}_{\rm xc}
  = \alpha \hat{v}_{\rm x}^{\rm hf} %%({\bf r},{\bf r^\prime}) 
  + (1-\alpha) \hat{v}_{\rm x}^{\rm gga} %%({\bf r})
  + \hat{v}_{\rm c}^{\rm gga} %% ({\bf r}) 
\end{equation}
where $\hat{v}_{\rm x}^{\rm gga}$ is the LDA/GGA exchange potential and $\hat{v}_{\rm c}^{\rm gga}$ is 
the LDA/GGA correlation potential. $\hat{v}_{\rm x}^{\rm hf}$ is the Hartree-Fock exchange potential which 
is a non-local ({\it i.e.} non-diagonal in real space) effective one-body potential:
\begin{equation}
  v_{\rm x}^{\rm hf}({\bf r},{\bf r^\prime}) = \bra{\bf r}\hat{v}_{\rm x}^{\rm hf}\ket{\bf r^\prime} 
= -\frac{1}{2}\rho({\bf r},{\bf r}^\prime)V_{ee}({\bf r}-{\bf r}^\prime)
\end{equation}
where $\rho({\bf r},{\bf r}^\prime)$ is the density matrix and 
$V_{ee}({\bf r}-{\bf r}^\prime)= \bra{{\bf r},{\bf r^\prime}} \hat{V}_{ee}\ket{{\bf r},{\bf r^\prime}} 
= 1/\|{\bf r}-{\bf r^\prime}\|$ is the bare Coulomb interaction between two electrons.

One can think loosely of the HYF approach as a first order correction of the KS Hamiltonian 
in a fraction $\alpha$ of the bare Coulomb interaction $\hat{V}_{ee}$.
This $\alpha\hat{V}_{ee}$ can be interpreted as a ``screened Coulomb interaction'' in a 
similar way as the $U$-parameter in the LDA+U method \cite{Anisimov:prb:91,Anisimov:jpcm:97a}. 
But in contrast to the LDA+U method, HYFs make the correction in the screenend 
Coulomb interaction on the {\it entire} LDA/GGA Hamiltonian and not only within a 
small subspace of atomic orbitals. Also note that the XC potential of the HYF method 
is non-local ({\it i.e.} dependent on both ${\bf r}$ and ${\bf r^\prime}$ instead of ${\bf r}$ alone) 
due to the contribution of exact Hartree-Fock exchange to the HYF. This shows that 
the HYF approach really falls outside the framework of conventional DFT where the XC potential 
is required to be local.

It turns out that the optimal amount $\alpha$ of Hartree-Fock exchange 
is almost universally of 20\%-25\% which has been rationalized by perturbation theory
\cite{Burke:jcp:96}. In the following we employ a functional similar to the popular B3LYP 
and B3PW functionals of quantum chemistry\cite{Becke:jcp:93b} that mix 20\% 
of Hartree-Fock exchange with 80\% of GGA exchange functionals but instead of GGA we employ plain LDA here
and refer to this HYF as LDA20. 

By adding the Hubbard term (\ref{eq:hubbard}) to the KS band structure 
some of the Coulomb interaction within the correlated subspace is double-counted 
since it has already been taken into account on a static mean-field level by the KS 
Hamiltonian $\hat{H}_{\rm ks}$. Thus the KS Hamiltonian has to be corrected by a
double-counting correction (DCC) term:
\begin{equation}
  \hat{H}({\bf k}) = \hat{H}_{\rm ks}({\bf k}) - \hat{h}_{\rm dc}
\end{equation}

In the case of LDA/GGA an exact expression for this DCC term $\hat{h}_{\rm dc}$ 
is not known, and several forms of the DCC have been suggested \cite{Petukhov:prb:03}. 
In the case of HYFs, however, at least the DCC for the Hartree-Fock contribution to 
the HYF is known exactly: It is the Hartree potential $\hat{v}_{\rm h}$ plus the 
Hartree-Fock exchange potential $\hat{v}_{\rm x}^{\rm hf}$ projected onto the correlated 
subspace. Below we argue that this is the only relevant contribution to the DCC term 
of the HYF+DMFT approach. However, we cannot proof this conjecture strictly so that
ultimately it must be justified by its success in the application to different 
materials.

Our argument goes as follows:
Bearing in mind that the HYF approach corresponds to a first order correction of the 
effective LDA Hamiltonian in a fraction $\alpha$ of the bare Coulomb interaction 
$\hat{V}_{ee}$, one first has to remove entirely this perturbation correction in 
$\alpha\hat{V}_{ee}$ in the correlated subspace since the DMFT calculation will 
treat the Coulomb interaction (locally) exact in that subspace. Thus if we take 
$\alpha V_{abcd}=\alpha\bra{ab}\hat{V}_{ee}\ket{cd}$ as the screened Coulomb interaction 
$U_{abcd}$ in the correlated subspace we obtain for the DCC of the HYF approach:
\begin{equation}
  \label{eq:hyfdcc}
  \bra{a\sigma}\hat{h}_{\rm dc}^{\rm hyf}\ket{b\sigma} 
  = \alpha \bra{a\sigma}\hat{v}_{\rm h}+\hat{v}_{\rm x}^{\rm hf}\ket{b\sigma} 
  + \bra{a\sigma}\hat{h}_{\rm dc}^{\rm ldac}\ket{b\sigma}
\end{equation}
where the last term $\bra{a\sigma}\hat{h}_{\rm dc}^{\rm ldac}\ket{b,\sigma}$ corrects
the double-counting in the LDA correlation potential $\hat{v}_{\rm c}^{\rm lda}$
within the correlated subspace. This term is not known exactly
since the contribution of a subspace to the total LDA correlation potential 
cannot be calculated exactly. However, the correlation potential is usually
much smaller than the exchange potential, and we will thus neglect this
contribution here. 

Assuming that only the direct Coulomb interactions $U=U_{abab}$ and the exchange Coulomb 
interactions $J=U_{abba}$ are important, we find the following simplified term for the 
on-site HYF DCC which allows us to predict the positions of the strongly correlated 
orbitals for the DMFT calculation:
\begin{equation}
  \label{eq:hyfdcc-simple}
  \bra{a\sigma}\hat{h}_{\rm dc}^{\rm hyf}\ket{a\sigma}
  \approx U( N_f - n_a^\sigma) - J( N_f^\sigma - n_a^\sigma)
\end{equation}
where $N_f$ is the total number of correlated electrons per atom and $n_a^\sigma$
is the number of electrons in atomic orbital $a$ with spin $\sigma$. 
Eq. (\ref{eq:hyfdcc-simple}) is different from the usual expression
for the LDA DCC in that the HYF DCC term becomes now orbital dependent
so that unoccupied orbitals experience a larger shift than occupied ones.
Also note that since the Hartree-Fock potential is non-local 
(i.e. non-diagonal in real space) the DCC term is also non-local.
But by construction the DCC term only acts on the correlated subspace. 

Within DMFT the self-energy is determined self-consistently by mapping 
the original problem onto an Anderson impurity problem. To this end the 
local Green's function,
\begin{equation}
  \label{eq:Gloc}
  \hat{G}_{\rm loc}(\omega) = \sum\nolimits_{\bf k} (\omega+\mu-\hat{H}({\bf k})-\hat\Sigma(\omega))^{-1}
 \end{equation}
projected onto the correlated subspace is equated to the Green's function of the
equivalent impurity problem:
\begin{equation}
  \label{eq:G0}
  \hat{G}_0(\omega) = (\omega+\mu-\hat{H}_0-\hat\Delta(\omega)-\hat\Sigma_f(\omega))^{-1}
\end{equation}
Here $\hat{H}_0$ is the (single-particle) Hamiltonian of the impurity site, 
$\hat\Delta(\omega)$ is the hybridization function with the conduction bath 
electrons, and $\hat\Sigma_f=\hat{P}_f\hat\Sigma\hat{P}_f$ is the full self-energy
projected onto the correlated subspace where $\hat{P}_f$ is the projection operator
for the correlated subspace.
The mapping $\hat{G}_0(\omega)\equiv \hat{P}_f\hat{G}_{\rm loc}(\omega)\hat{P}_f$ defines the so-called 
self-consistency condition which is central to the DMFT method. 
The mapping yields the hybridzation function,
\begin{equation}
  \label{eq:Delta}
  \hat\Delta(\omega) = \omega+\mu - \hat{H}_0 - \hat\Sigma_f(\omega) 
  - (\hat{P}_f\hat{G}_{\rm loc}(\omega)\hat{P}_f)^{-1}
\end{equation}
with $\hat{H}_0\equiv\sum_{\bf k} \hat{P}_f\hat{H}({\bf k})\hat{P}_f$.
These are the relevant quantities for solving the impurity problem. 
By solving the impurity problem, one obtains in turn the self-energy 
$\hat\Sigma(\omega)$. Equations (\ref{eq:Gloc})-(\ref{eq:Delta}) 
define the self-consistent DMFT procedure for computing the self-energy 
$\hat\Sigma(\omega)$.

Solving the impurity problem given by ($\ref{eq:G0}$) is the computationally 
most demanding step in most DMFT calculations. A variety of solvers ---each 
suitable for a certain region of parameters--- is available to deal with the 
impurity problem. An overview over the different techniques can be found in {\it e.g.}
Ref. \cite{Kotliar:rmp:06}. Since we are interested in describing a Mott 
insulator a suitable method for solving the impurity problem 
is an expansion in the hybridization strength in the so called non-crossing 
approximation (NCA) \cite{Haule:prb:01}.

\section{Results}
In order to show how the HYF method efficiently improves
the gap of band insulators in comparison with conventional DFT methods
we first perform LDA and HYF calculations of La$_2$O$_3$ which is similar 
to Ce$_2$O$_3$ but does not have the strongly correlated $4f$-electrons.
It is a typical band insulator with a reported band gap of about 
5.5~eV between the O $2p$-valence band and the La $5d$-conduction band 
\cite{Adachi:chemrev:98}. 
For the LDA and HYF calculations we employ here and in the following 
LDA+DMFT and HYF+DMFT calculations of Ce$_2$O$_3$
the CRYSTAL06 {\it ab initio} electronic structure program for crystalline solids 
\cite{Crystal:06} together with a Gaussian basis set and pseudo potential
by Cundari and Stevens tailored for the rare earth series elements 
\cite{Cundari:jcp:93}. The basis set retains the $4f$, $5s$, $5p$, $5d$, $6s$ and $6p$ 
shells for the electronic structure calculations while the inner shells are described 
by an effective core pseudo potential.
In the self-consistent LDA and HYF calculations the irreducible Brillouin zone is sampled
by 193 $k$-points while in the DMFT calculations the entire Brillouin zone is sampled by
1000 $k$-points. The DMFT calculations are performed at a finite 
temperature of $k_BT=0.2$~eV. The LDA+DMFT and HYF+DMFT calculations are not
fully self-consistent in the charge density\cite{Pourovskii:prb:07}.

\begin{figure}
  \includegraphics[width=0.9\linewidth]{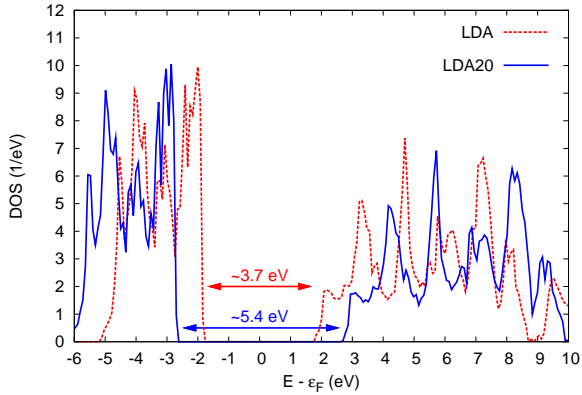}
  \caption{(Color online)
    DOS of La$_2$O$_3$ calculated with an LDA functional 
    (dashed red line) and with the LDA20 HYF
    (continuous blue line). 
  }
  \label{fig:la2o3}
\end{figure}

Fig. \ref{fig:la2o3} shows a comparison between the density of states 
(DOS) calculated with a pure LDA functional on the one hand and the LDA20
HYF on the other hand. While in the LDA calculation 
the magnitude of the band gap is underestimated as expected by more than 1.5~eV, 
the HYF calculation does indeed give the experimental value of the band gap 
of about 5.5~eV. 

We would like to stress here that changing the LDA or GGA functional part of the
HYF does not alter the results significantly. In fact, we have obtained
very similar results and the correct magnitude of the band gap also with the popular 
B3LYP or B3PW functionals. In contrast the results depend quite strongly on the 
exact amount of Hartree-Fock exchange. Thus increasing the amount of Hartree-Fock exchange 
to 25\% already increases the magnitude of the band gap to about 6~eV.

To demonstrate the above developed HYF+DMFT method we apply it now to
the exemplary case of Ce$_2$O$_3$. As before we employ the LDA20 HYF
which reproduces correctly the band gap of La$_2$O$_3$.  The starting 
point is a HYF calculation of the paramagnetic phase of Ce$_2$O$_3$
which results in a metallic state with the Fermi level in the $4f$-band. 
In order to prevent excessive symmetry breaking of the Ce $4f$-orbitals 
prior to the DMFT calculations, the calculation is done at a finite 
temperature of $k_BT\approx0.5$~eV.

In spite of the quite high temperature and although the crystal field 
splitting is actually quite weak (of order 0.1~eV), the \emph{orbital} 
symmetry of the $4f$-orbitals has been broken: The energy difference 
between lowest and highest $4f$-orbital is about 1.2~eV. 
The reason for the symmetry-breaking is the Hartree-Fock contribution to the HYF
which tends to break symmetries to lower the energy of the system so that 
those $4f$-orbitals that are slightly favored by the weak crystal field splitting 
become more occupied during the self-consistent solution of the KS equations while 
those unfavored by the crystal field splitting become less occupied. Therefore the 
splitting is strongly enhanced by the Hartree-Fock term. 
Since the Hartree-Fock term is the principal responsible for the symmetry-breaking
the energy difference between lowest and highest $4f$-orbital is reduced to 0.4~eV
when applying the HYF DCC scheme outlined above, partially restoring the symmetry.
This can be understood by considering the orbital dependence of the HYF 
DCC (\ref{eq:hyfdcc-simple}) which shifts occupied orbitals less than unoccupied
orbitals decreasing the symmetry-breaking. At lower temperatures the symmetry-breaking 
becomes even stronger: E.g. for $k_BT\approx0.2$~eV the splitting of the $4f$-orbitals
is still of about 2.3~eV \emph{after} applying the DCC. 

We note that the on-site Coulomb repulsion $\alpha V_{aaaa}$ corresponding to the 
Hartree-Fock contribution is about 5~eV. This value is only slightly smaller than the $U$ 
usually employed for the Ce $4f$-orbitals in actual LDA+U an DMFT calculations 
\cite{Pourovskii:prb:07,Loschen:prb:07,Andersson:prb:07} which is between 5.5~eV and 6.5~eV.
This again points to the correctness of the interpretation of $\alpha V_{aaaa}$ as a 
screened Coulomb interaction similar to the $U$ of the LDA+U method \cite{Anisimov:prb:91}.

\begin{figure}
  \includegraphics[width=0.9\linewidth]{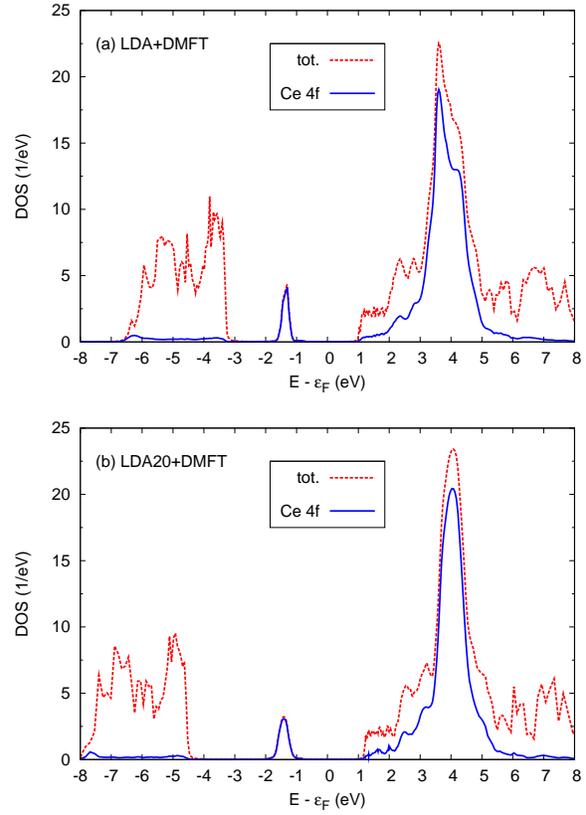}
  \caption{(Color online)
    Tot. DOS (dashed red line) and partial Ce $4f$ DOS (continuous blue line) 
    of Ce$_2$O$_3$ calculated with LDA+DMFT (a) compared to the 
    HYF+DMFT approach (b) explained in the text. The on-site Coulomb 
    repulsion for the DMFT calculation is $U=5$~eV and $J=0.2$~eV in both cases. 
  }
  \label{fig:ce2o3:dft+dmft}
\end{figure}

In Fig. \ref{fig:ce2o3:dft+dmft} we compare DMFT calculations (a) on top of the
plain LDA calculation and (b) on top of the LDA20 HYF calculations. In both cases the 
$4f$-band splits into a lower Hubbard band filled with one electron per Ce atom and an 
upper unoccupied Hubbard band. We take $\alpha{}V_{aaaa}=5$~eV as an estimate for $U$
and similarly $\alpha V_{abba}=0.2$~eV as an estimate for $J$. The resulting band gap 
between the occupied $4f$-Hubbard band and the empty $5d$-conduction band is about 2.3~eV 
in both cases which is in very good agreement with the measured band gap of 2.5~eV 
\cite{Adachi:chemrev:98}. The agreement can be improved by slightly increasing $U$ 
to about 5.5~eV. Most importantly, the $pd$-gap between the occupied O $2p$-bands and 
the unoccupied Ce $5d$-bands is 5.5~eV in the HYF+DMFT calculation and is thus exactly 
the experimentally measured $pd$-gap of about 5.5~eV. This is a considerable improvement 
over the LDA+DMFT where the $pd$-gap is only of about 4~eV. 
The occupancy $N_f$ of the Ce $4f$-shell is very similar in both methods:
We find $N_f=1.00$ for the LDA+DMFT, and $N_f=0.98$ for the HYF+DMFT 
calculation. Finally, we mention that we have also obtained satisfactory results
(not shown) for the Nd$_2$O$_3$ compound using the same methodology. 

\section{Conclusions}
In conclusion, we have proposed a new method that combines the HYF
approach with DMFT. We have shown that this HYF+DMFT method gives a qualitatively 
\emph{and} quantitatively correct description of the electronic structure 
of Ce$_2$O$_3$ as a prototypical example of the strongly correlated insulating Rare 
Earth compounds. While the HYF part fixes the magnitude of the $pd$-band gap which is 
underestimated in the LDA+DMFT approach, the DMFT part takes care of the strongly 
localized $4f$-electrons which are not properly taken into account within conventional 
KS band theory. It thus predicts correctly the opening of the Mott-Hubbard gap 
in the Ce $4f$-band in addition to the band gap between the O $2p$-valence band and the 
Ce $5d$-conduction band. While the HYF is computationally slightly more expensive than 
the LDA, its cost is either comparable or less than that of the DMFT calculation.
It is also computationally considerably less demanding than the GW 
approximation which gives an accurate description of the quasi-particle spectra of 
weakly correlated materials \cite{Hedin:pr:65,Kotani:prb:07} and can be combined with 
DMFT in a natural way \cite{Biermann:prl:03}. Hence, the HYF+DMFT approach is an attractive 
avenue for improving the overall accuracy of spectra in materials containing both correlated 
and uncorrelated electrons at reasonable computational cost.

\acknowledgments
We would like to thank Antoine Georges for suggesting the rare earth oxide 
problem to us and for numerous insightful comments. 
We thank Roberto Dovesi for useful hints on running the CRYSTAL06 program
and Silke Biermann, Victor Oudovenko, Leonid Pourovski, Erick Wimmer, 
and Rene Windiks for fruitful discussions. 
DJ and GK acknowledge funding by the NSF under grant No. DMR 0528969.

\bibliographystyle{eplbib}
\bibliography{materials,theory}

\end{document}